\documentclass{elsart3p}

\usepackage{amssymb}
\usepackage{graphicx}

\begin{document}

\begin{frontmatter}

%titles, authors and addresses


\title{Electric and magnetic characterization of NbSe$_2$ single crystals: 
anisotropic superconducting fluctuations above $T_C$
}


\author[a]{F. Soto},
\author[b]{H. Berger},
\author[a]{L. Cabo},
\author[a]{C. Carballeira},
\author[a]{J. Mosqueira\corauthref{cor1}},
\corauth[cor1]{Corresponding author. e-mail: fmjesus@usc.es}
\author[b]{D. Pavuna},
\author[a]{P. Toimil},
\author[a]{F. Vidal}

\address[a]{LBTS, Departamento de F\'isica da Materia Condensada, Universidade de Santiago de Compostela, E-15782 Spain}
\address[b]{Department of Physics, Ecole Politechnique F\'ed\'erale de Lausanne, CH-01015, Lausanne, Switzerland}

\begin{abstract}
Electric and magnetic characterization of NbSe$_2$ single crystals is first presented in detail. Then, some preliminary measurements of the fluctuation-diamagnetism (FD) above the transition temperature $T_C$ are presented. The moderate uniaxial anisotropy of this compound allowed us to observe the fluctuation effects for magnetic fields $H$ applied in the two main crystallographic orientations. The superconducting parameters resulting from the characterization suggest that it is possible to do a reliable analysis of the FD in terms of the Ginzburg-Landau (GL) theory.
\end{abstract}

\end{frontmatter}

% main text
%\section{Title of the first section}
%\label{labelOfFirstSection}
The anisotropy of the FD above $T_C$ is still a relatively unexplored issue. The GL theory predicts that the fluctuation-induced magnetic susceptibility when $H$ is parallel to the crystallographic $ab$ planes is related to the one when $H\perp   ab$ through $\Delta\chi_\parallel=\Delta\chi_\perp/\gamma^2$, where $\gamma$ is the anisotropy factor.[1]  While most of low-$T_C$ superconductors are isotropic, the high $T_C$'s are so anisotropic that $\Delta\chi_\parallel$ is practically unobservable (for the less anisotropic one, YBa$_2$Cu$_3$O$_7$ with $\gamma \approx10$, $\Delta\chi_\parallel\approx10^{-2}\Delta\chi_\perp$). The moderate anisotropy of NbSe$_2$ ($\gamma\approx 3$, and then $\Delta\chi_\parallel\stackrel{>}{_\sim}10^{-1}\Delta\chi_\perp$) makes this material an excellent test bed to explore anisotropy effects on the FD above $T_C$. Here we present a magnetic and electric characterization of NbSe$_2$ single crystals. The resulting superconducting parameters will allow a reliable analysis of the FD measured above $T_C$. 

\begin{figure}
\begin{center}
\includegraphics*[width=6cm]{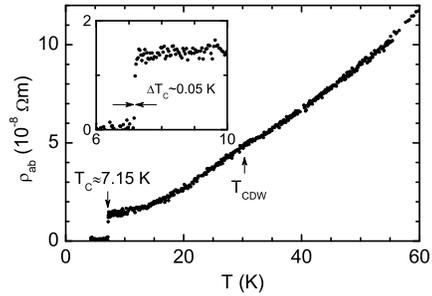}
\end{center}
\caption{Temperature dependence of the in-plane electrical resistivity. The kink at $\sim30$ K corresponds to a \textit{charge-density wave} (CDW) transition.}
\end{figure}

\begin{figure}[t]
\begin{center}
\includegraphics*[width=6cm]{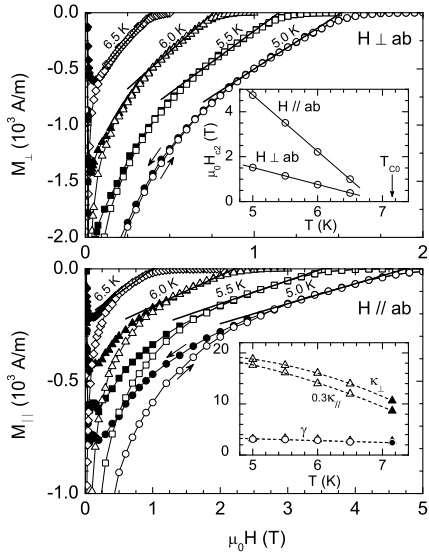}
\end{center}
\caption{$M(H)$ isotherms below $T_C$. The fits of the Abrikosov theory for the mixed state (thick lines) leads to the $H_{C2}^{\perp,\parallel}$,  $\kappa_{\perp,\parallel}$ and $\gamma$  values presented in the insets.}
\end{figure}

The NbSe$_2$ single crystals used in this work were synthesized by chemical transport method. The transition temperature $T_C$ = 7.15 K was determined from the temperature dependence of the in-plane electrical resistivity $\rho_{ab}$ (Fig. 1). This measurement was done in a $2\times1\times0.067$ mm$^3$ single crystal cut from one of the crystals used in the magnetization measurements, by using a \textit{van der Pauw} contacts configuration. The ratio $\rho_{ab}$(300 K)/$\rho_{ab}$(7.5 K) = 43, is comparable to the ones found in the best crystals.[2] The in-plane mean free path  $\ell_{ab}\approx$ 1830 \r{A} was obtained from the in-plane residual resistivity and from the carriers concentration,[3] by using a Drude-model relation. The resistivity anisotropy is found to be $\rho_{c}/\rho_{ab}\sim 60$, which leads to a mean free path in the $c$ direction of $\ell_{c}\approx$ 30 \r{A}. The magnetization measurements were performed with a commercial (Quantum Design) SQUID magnetometer. In Fig. 2 we present $M(H)$ measurements at different temperatures below $T_C$ and with $H\perp ab$ and $H\parallel ab$. The thick lines in this figure are the best fit to the data points of the high-field mixed state magnetization of anisotropic superconductors [3]:
\begin{equation}
M_{\perp,\parallel}(T,H)\approx[H-H_{C2}^{\perp,\parallel}(T)]/2\beta_A\kappa_{\perp,\parallel}
\end{equation}
where $H_{C2}^{\perp,\parallel}$ are the upper critical magnetic fields for $H\perp ab$ and $H\parallel ab$  respectively, $\kappa_\perp=\lambda_{ab}/\xi_{ab}$ and $\kappa_\parallel=\gamma\kappa_\perp$ are the corresponding GL parameters ($\lambda_{ab}$ and $\xi_{ab}$ are, respectively, the in-plane magnetic penetration and coherence lengths) and  $\beta_A\approx1.16$. The resulting $H_{C2}^{\perp,\parallel}$ and $\kappa_{\perp,\parallel}$  are presented in the insets of Fig. 2. As may be clearly seen, both upper critical fields are linear in $T$ in the temperature range studied, and extrapolate to $H_{C2}^{\perp}(0)\approx 5.3$ T and $H_{C2}^{\parallel}(0)\approx 17.3$ T. By combining these values with the GL expressions $\mu_0H_{C2}^{\perp}=\phi_0/2\pi\xi_{ab}^2$ and $\mu_0H_{C2}^{\parallel}=\phi_0/2\pi\xi_{ab}\xi_c$, we obtained $\xi_{ab}(0) = 78.8$ \r{A} and  $\xi_{c}(0) = 24.1$ \r{A}. The GL parameters extrapolate to $\kappa_\perp\sim11$ and respectively $\kappa_\parallel\sim 29$ at $T_C$. The anisotropy factor as calculated from  $H_{C2}^{\parallel}/H_{C2}^{\perp}$ and from $\kappa_{\parallel}/\kappa_{\perp}$ tends to $\sim2.4$ at $T_C$. The agreement of both  $\gamma$  determinations is an important consistency check of the procedure used to obtain the NbSe$_2$ superconducting parameters, which also agree with the values in the literature.[4]

A first example of the as-measured $M(T)$ above $T_C$ for $H\perp ab$ is presented in Fig. 3. The dashed line is the normal-state contribution, which was obtained by fitting a Curie-like function well above $T_C$ (for $T > 1.5 T_C$). The solid line is the prediction of the anisotropic 3D-GL theory under a total-energy cutoff (see, e.g., Ref. [5]) evaluated with the above superconducting parameters. As may be clearly seen, the theory overestimates the experimental FD by a factor of $\sim2$, which cannot be explained by the background determination uncertainty. On the contrary, when $H \parallel ab$ we found an excellent agreement between the theory and the experimental results. We suggest that these results may be explained in terms of the anisotropy of non-local electrodynamic effects on the FD. Also, the FD seems not to be appreciably affected by the CDW state appearing below $\sim35$ K.

\begin{figure}[t]
\begin{center}
\includegraphics*[width=6cm]{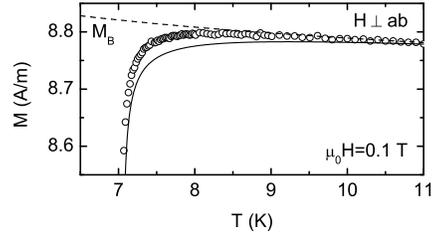}
\end{center}
\caption{$M(T)$ curve above $T_C$ for $H\perp ab$. $M_B$ is the normal-state contribution, and the solid line is the prediction of the anisotropic 3D-GL theory. }
\end{figure}

Supported by MEC and FEDER funds (MAT2004-04364), and by Xunta de Galicia (PGIDIT04TMT 206002PR). HB and DP acknowledge support by the EPFL and the Swiss National Science Foundation, and LC by a MEC FPU grant.

\end{document}